\documentstyle[natbib209,aasj,mnfixes,epsfig]{mn-nat}
\begin{document}

\title[Starbursts near black holes]
{Starbursts near supermassive black holes: young stars in the Galactic Center, and gravitational waves in LISA band.}

\author[Levin ]{Yuri Levin$^{1,2}$ 
\\
$^1$Leiden Observatory, P.O. Box 9513, NL-2300 RA Leiden
\\
$^2$Lorentz Institute, P.O. Box 9506, NL-2300 RA Leiden}

\date{printed \today}
\maketitle
\begin{abstract}
We propose a scenario in which massive stars
form in a self-gravitating gaseous disc around
a supermassive black hole.
We analyze the dynamics of a
disc forming around a supermassive black hole,
in which the angular momentum   is
transported by  turbulence induced
by the disc's self-gravity. 
We find that once the surface density of
the disc exceeds 
a critical value, 
the disc fragments into dense clumps. 
 We argue that the clumps accrete material from the remaining
disc and merge into larger 
clumps;
the upper mass of a merged clump is a few tens to
a few hundreds of solar mass. 

This picture fits well with the observed 
young stellar discs near the SgrA* black hole in the Galactic Center. In particular, we show how the masses and
spatial distribution of the young stars, and the total mass in the Galactic Center 
discs can
be explained.
However,
explaining the origin of the several young stars closest to the black hole (the S-stars) is more problematic: their orbits are compact,
eccentric, and have random orientation. We propose that the S-stars were
born in a previous starburst(s), and then migrated through their parent
disc via type I or runaway migration. Their orbits were then randomized by the Rauch-Tremaine
resonant relaxation.

We then explore the consequences of the star-formation scenario
for AGN discs, which are continuously resupplied with gas. 
We argue that some compact remnants generated by the starburst will get embedded
in the disc. 
The disc-born stellar-mass black holes will
interact gravitationally with the massive accretion disc
and be dragged
towards the central black
hole. Merger of a disc-born
black hole  with the central black hole will produce a burst
of gravitational waves. If the central black hole 
is accreting at a
rate comparable to the Eddington limit, the gas drag from the 
accretion disc will not alter significantly the dynamics
of the final year of merger, and the gravitational
waves should be observable  
by LISA. For a reasonable range of parameters
 such mergers will be detected
monthly, and that the gravitational-wave signal from these mergers is
distinct from that of other merger scenarios. Also, for some plausible 
black hole masses and accretion rates, the burst of gravitational waves
should  be accompanied by a detectable change in the  optical luminosity
of the central engine.

\end{abstract}

\section{introduction}

It is widely believed that self-gravitating accretion discs can form 
around supermassive black holes (SBH) in AGNs.
Theoretical models show that the AGN  accretion discs must become
self-gravitating if they extend beyond a fraction of a parsec
away from a central black hole (Paczynski 1978, Kolykhalov and
Sunyaev 1980, Schlosman and Begelman 1987,
 Kumar 1999, Jure 1999, Goodman 2003). Self-gravitating discs are
unstable to fragmentation on a dynamical timescale; 
self-gravity of AGN accretion discs is a major 
issue in understanding how gas is delivered 
to the central black hole. It is likely that star formation
will occur in the outer parts of an AGN
  accretion disc (Kolykhalov and Sunyaev 1980, 
Schlosman and Begelman 1987).

There are two lines of observational 
evidence that a starburst within SBH's radius of influence may be a common phenomenon:

(i) Levin and Beloborodov (LB, 2003), using pre-existing data, have identified a 
disc of young massive stars which are moving clockwise in the gravitational potential of SgrA*, 
the SBH at the center of the Galaxy. They have argued that the stars were born in a dense accretion disc 
which existed several million years ago in the Galactic Center. The presence of the disc was later confirmed by 
Genzel et. al. (2003) and Paumard et al. (2005), who have used updated data sets; 
these authors also found a second 
counter-clockwise disc of young massive stars. For the past few years, an alternative possibility for the 
origin of the young stars was considered on equal footing with LB'S proposal. In the alternative scenario 
(Gerhard 2001, Hansen and Milosavljevic 2003),
 the stars were born tens of parsecs from SgrA* and were originally members of a massive star 
cluster. This cluster then spiraled in towards SgrA* due to dynamical friction with an inner bulge, and deposited 
the stars 
near SgrA*.  However, this scenario is now disfavored due to two recent observations: (a) In the cluster scenario 
one expects many young low-mass stars to also be present in the Galactic Center and for them to produce copious 
amounts of x-ray emission from their coronae. However, Chandra observations show that the Galactic Center is a 
relatively 
weak x-ray emitter, and thus the presence of a multitude of low-mass stars is ruled out 
(Nayakshin and Sunyaev, 2005); (b) Paumard et al. (2005) find no early-type stars outside of the central 
half-parsec, which 
again strongly argues against the in-spiraling cluster scenario. Therefore, most likely the young stellar disc 
formed in situ as result of fragmentation of a gaseous disc, as was argued by LB.

(ii) Bender et al.~(2005) have used HST to identify a compact disc of A-stars which is located deep 
in the gravitational potential of a SBH in the nucleus of M31. The stars are likely much younger 
than the SBH (this is true unless one assumes that the SBH is older than $\sim 10^8$ years),
 and the strong tidal barrier makes it unlikely that the disc is a remnant of a tidally disrupted 
star cluster since any cluster would get disrupted at a larger distance from the SBH (Nayakshin, 2005). 
Therefore, the disc of A-stars is most likely the remnant of a gaseous accretion disc which 
existed in M31 about a hundred million years ago.

Star formation in an SBH radius of influence may be connected to supporting the high accretion of some AGNs (Thompson et al.~2005)
and at the same time  may help explain low luminocity of others (Tan and  Blackman 2005).
The dynamics of the fragmenting disc is strongly affected by the feedback energy input
from the starburst.
Collin and Zahn (1999) have conjectured that the feedback from this star formation 
will prevent the accretion disc from
becoming strongly self-gravitating. However,
Goodman (2003) has used general energy arguments to show
that the feedback from star formation is insufficient to
prevent an AGN disc  with the near-Eddington accretion rate
from becoming strongly self-gravitating
at a distance of $10^4$---$10^5$ Scwartzschild radii from the central black hole (about $0.1$pc
for our Galactic Center). This is
distinct from the case of galactic gas discs, for which  there
is evidence that the feedback from star formation protects  them  from
their self-gravity. 

In this paper we concentrate
on the physics of  the  self-gravitating disc and
make a semi-analytical estimate  of the possible
mass range of  stars formed in such  discs (Sections 2
and 3). 
 Our principal
conclusion is that the stars can be very massive,
up to hundreds of solar masses. This conclusion is in qualitative agreement
with two recent independent observations: Nayakshin and Sunyaev (2005) and Paumard et al.~(2005)
have shown that the young stars in the GC must were produced in a  starburst with the top-heavy IMF
strongly favoring massive stars.  
In Section 4 we specialize to the case of SgrA* discs and show that the mass and the column density distribution
of the marginally fragmenting disc are consistent with those  of the currently observed stellar discs.
We also address the puzzle of the several the young stars in the central arc-second (the S-stars). Their
orbits present a problem for the disc-starburst
picture. Their extreme proximity to SgrA*, eccentric orbits, and random inclinations exclude the possibility
that they were born from the gaseous disc at their current location. We argue that instead they 
were born in a disc at a larger
distance from SgrA*, but then migrated inwards due to gravitational interaction
with the disc. Their eccentricity and inclination 
angles were randomized by relatively fast resonant relaxation, a process discovered by Rauch and Tremaine in
1996.

In section 5,  we consider a self-gravitating AGN accretion disc
which is continuously supplied by gas on a timescale greater than the lifetime of
massive stars. Some of the black-hole remnants of the stars become embedded in the
disc and  
and migrate inward 
on the timescale of $\sim 10^7$ years.  
The merger of the migrating black hole with the central
black hole will produce gravitational waves.
We show that for a broad range of
AGN accretion rates the final inspiral
is unaffected by gas drag, and therefore the
gravitational-wave signal should be detectable
by LISA.
 The  rate
of these mergers is uncertain, but if even a fraction of
a percent of the disc mass is converted into black holes
which later merge with the central black hole, then LISA should
detect monthly a signal from such a merger. The
final inspiral may occur close to the equatorial plane
of the central supermassive hole and is likely
to follow a quasi-circular orbit, which would make the
gravitational-wave signal distinct from those
in  other astrophysical merger scenarios.
If the disc-born black hole is sufficiently
massive, it will disrupt accretion flow in the disc
during the final year of its inspiral, thus making
an optical counterpart
to the gravitational-wave signal.

\section{Physics of a fragmenting self-gravitating disc}
The importance of self-gravitating accretion discs
in astrophysics  has long been understood (Paczynski 1978,
Lin and Pringle 1987). It was conjectured that the turbulence
generated by the gravitational (Toomre) instability may act as a 
source of viscosity in the disc. This viscosity  would both
drive accretion and keep the disc hot; the latter would act as
a  negative feedback for the Toomre instability
and would keep  the disc only marginally unstable.
Recently, there has been  big progress in our understanding
of the self-gravitating discs, due to a range of new and sophisticated
numerical simulations (Gammie 2001, Mayer et.~al.~2002,
Rice et.~al.~2003).
In our analysis, we shall rely extensively on these numerical
 results.

Consider an accretion disc 
which  is  supplied
by a gas infall. 
This  situation may arise when a merger or some other
major event in a  galaxy  delivers gas to the proximity of a supermassive
black hole residing in the galactic bulge. 
Let $\Sigma (r)$ be the surface density of the disc.
We follow the evolution of the disc
as $\Sigma(r)$ gradually increases due
to the  infall.

We begin by assuming that initially there is no 
viscosity mechanism, like Magneto-Rotational Instability
(MRI), to transport
the angular momentum and keep the disc hot\footnote{When
the disc begins to fragment, the viscosity due to self-gravity-driven
shocks exceeds the one due to MRI; see below. Therefore, while computing
the disc parameters at fragmentation, it is reasonable to ignore MRI}. 
This assumption
is valid when the ionization fraction of the gas in the
disc is low, i.e.~when the gas is 
far enough from a central source (about a thousand 
Schwarzschild radii from the supermassive black hole).
 We
also, for the time being, neglect irradiation from the central
source; this may be a good assumption if
a disrupted molecular cloud forms a disc
but the accretion onto the hole
has not yet begun.
 As will be discussed below,
irradiation is important for some regions
in the AGN discs we are considering. However,
as shown in the following subsection,
inclusion of irradiation or other
source of heating will only
strengthen the case for formation of massive stars.
\begin{figure}
\begin{center}
%\hbox{
\epsfig{file=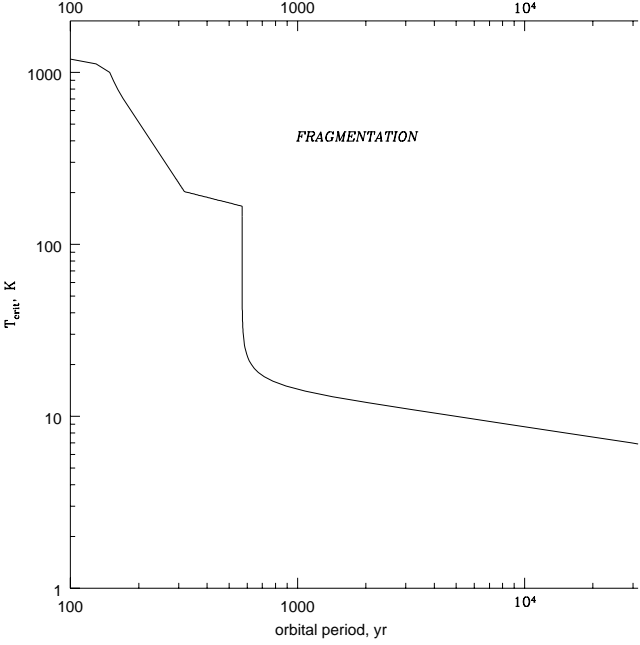, width=9cm}
%}
\end{center}
\caption{The temperature of the critically fragmenting disc as a function
of the orbital period.}
\end{figure}

We assume,
 therefore, that the forming disc  cools until
it becomes self-gravitating; this happens
when
\begin{equation}
Q={c_s\Omega\over \pi G \Sigma}\simeq 1.
\label{Q}
\end{equation}
Here $c_s$ is the isothermal speed of sound at the mid-plane
of the disc, and $\Omega$ is
the angular velocity of the disc.
Numerical simulations show that once the disc becomes self-gravitating,
turbulence and shocks develop; they
transport angular momentum and provide
heating which compensates the cooling of the disc
(Gammie 2001). We thus assume, in agreement with the simulations,
that when the disc exists, it is marginally
self-gravitating, i.e.~ $Q=1$. Then 
\begin{equation}
c_s={\pi G\Sigma\over \Omega},
\label{c_s}
\end{equation}
and
\begin{equation}
T\sim 2m_p c_s^2/k_B={2m_p\over k_B}\left(
   {\pi G\Sigma\over \Omega}\right)^2.
\label{T}
\end{equation}
Here  $T$ is the temperature in the
mid-plane of the disc, $m_p$ is the proton mass, and $k_B$
is the Boltzmann constant.

The one-sided flux from the disc
surface is given by the  modified Stephan-Boltzmann
law:

\begin{equation}
F=\sigma T_{\rm eff}^4,
\label{flux1}
\end{equation}
where $T_{\rm eff}$ is the effective temperature. It is related
to the mid-plane temperature by
\begin{equation}
T_{\rm eff}^4\sim T^4  f(\tau)=f(\tau)\left({2m_p\over k_B}\right)^4
       \left(
   {\pi G\Sigma\over \Omega}\right)^8.,
\label{Trelation}
\end{equation}
where $\tau=\kappa \Sigma/2$ is the optical
depth of the disc; here $\kappa(T)$ is the opacity of the disc.
The function $f(\tau)=\tau$ for an optically thin disc, and
$f(\tau)=1/\tau$ for an optically thick disc.
We combine these two cases in our model by taking
\begin{equation} 
f(\tau)={\tau\over \tau^2+1}.
\label{ftau}
\end{equation}
We have used Eq.~(\ref{T}) in the last step of Eq.~(\ref{Trelation}).
The flux from the disc is powered by the accretion energy:
\begin{equation}
F={3\over 8\pi}\Omega^2 \dot{M}={9\over 8}\alpha\Omega c_s^2 \Sigma=
  {9\over 8}\alpha (\pi G)^2 {\Sigma^3\over \Omega}.
\label{flux2}
\end{equation}
Here $\alpha$ parametrizes viscous dissipation due to 
self-gravity (Gammie 2001); we have used Eq.~(\ref{c_s}) in the last step.
Using Eqs (\ref{flux1}), (\ref{flux2}), and (\ref{Trelation}), we can
express the viscosity parameter $\alpha$ as
\begin{equation}
\alpha={8\sigma \over 9}\left({2m_p\over k_B}\right)^4(\pi G)^6
       f(\tau){\Sigma^5\over \Omega^7}.
\label{alpha1}
\end{equation}
It is very important to emphasize that in this model
$\alpha$ is only a function of $\Sigma$ and $\Omega$:
the temperature in the mid-plane
is determined by Eq.~(\ref{T}), and this temperature
sets the opacity which in turn determines the
optical depth $\tau=\kappa \Sigma/2$. The  opacity
in the range of temperatures and densities
of interest to us is set by light scattering
off ice grains and, in some cases, by scattering off
metal dust. The relevant regimes are worked  out in the literature
on protoplanetary discs; we use the analytical fit
to the opacity (in cm$^2$/gm) from  the appendix of Bell and Lin, 1994.
\begin{eqnarray}
\kappa=0.0002\times T_K^2&\hskip 0.2in\hbox{for}&\hskip .2in T<166\hbox{K},\nonumber\\
\kappa=2\times 10^{-16}T_K^{-7}&\hskip 0.2in\hbox{for}&\hskip .2in166\hbox{K}<T<202\hbox{K}
\label{iceopacity}\\
\kappa=0.1\sqrt{T_K}&\hskip .2in\hbox{for}&\hskip .2in T>202\hbox{K}.\nonumber
\end{eqnarray}
In the first interval the opacity is due to ice grains; in the
second interval the ice grains evaporate, and the opacity
drops sharply with the temperature; in the third interval
(the highest $T$)  dust grains are the major source of the opacity.

From Eq.~(\ref{alpha1}) we see that
 as $\Sigma$ of the disc increases
due to the merger-driven infall, the effective viscosity will
reach $\alpha_{\rm crit}\sim 1$. At this stage the cooling time of the
disc becomes comparable to the orbital period. Gammie's   simulations
show that in this case the turbulence induced by self-gravity
is no longer able to keep the disc together, and the disc fragments.
Gammie's simulations give $\alpha_{\rm crit}\simeq 0.3$; the numerical
value we quote  disagrees with Gammie's,
but agrees with $\alpha_{\rm crit}$ quoted by  Goodman (2003) since like Goodman
we use isothermal speed of sound for $\alpha$-prescription.
Gammie's results, although obtained
for razor-thin discs, justify the key assumption of our model:
{\it the disc exists as a whole for $\alpha<\alpha_{\rm crit}$, and fragments
once $\alpha=\alpha_{\rm crit}$}.  Similar criterion
for the disc fragmentation was already used in Shlosman and
Begelman, 1987. 

We shall refer to the mid-plane
temperature and surface density of the marginally
fragmenting disc with $\alpha=\alpha_{\rm crit}$ as the critical
temperature $T_{\rm crit}$ and the critical surface density,
$\Sigma_{\rm crit}$.

We now find the critical surface density and mid-plane temperature
as a function of $\Omega$. We use Eq.~(\ref{T}) to express
$\Sigma$ as a function of $T$ and $\Omega$,  then substitute
this
function into Eqs.~(\ref{ftau}) and (\ref{alpha1}), and set
$\alpha=\alpha_{\rm crit}$. After simple algebra,
we obtain
\begin{equation}
\Omega^3+p\Omega=q,
\label{omega1}
\end{equation}
where
\begin{eqnarray}
p&=&2\left({\pi G\over \kappa(T_{\rm crit})}\right)^2{m_p\over k_B T_{\rm crit}},\nonumber\\
q&=&{32\sigma\over 9\alpha_{\rm crit}}{(\pi G)^2\over \kappa(T_{\rm crit})}
      (m_p/k_B)^2 T_{\rm crit}^2.\label{pq}
\end{eqnarray}

There is an analytical solution to Eq.~(\ref{omega1}):
\begin{equation}
\Omega=w-p/(3w),
\label{omega2}
\end{equation}
where
\begin{equation}
w=\{q/2+[(q/2)^2+(p/3)^3]^{1/2}\}^{1/3}.
\label{w}
\end{equation}

In Figure $1$ we make a plot of $T_{\rm crit}$
as a function of the orbital period, for concreteness
we set $\alpha_{\rm crit}=0.3$. The critically
self-gravitating disc is optically thin
if the second term of the LHS of Eq.~(\ref{omega1})
is dominant, and optically thick otherwise. This
can be expressed as a condition on the critical
temperature: the disc is optically thin if
\begin{equation}
T_{\rm crit}<12\hbox{K}(\alpha_{\rm crit}/0.3)^{2/15},
\label{tcrit1}
\end{equation}
and optically thick for higher critical temperatures.
The angular frequency above which the critically 
unfragmented disc becomes optically thick is 
\begin{equation}
\Omega_{\rm transition}\simeq 16.3\times 10^{-11}\hbox{sec}^{-1}.
\label{omegatrans}
\end{equation}

We use Eqs.~(\ref{c_s}) and (\ref{T}) to find the
critical surface density $\Sigma_{\rm crit}$, which is plotted
in Fig.~2,
and the scaleheight $h_{\rm crit}=c_s/\Omega$ 
of a marginally fragmenting disc. 
The Toomre mass $\bar{M}_{\rm cl}=\Sigma_{\rm crit}
h_{\rm crit}^2$ is the mass scale of the first clumps which
form in the first stage of fragmentation. In Fig.~3, we plot the Toomre
mass of the critically fragmenting disc as a function
of the orbital period.

The value of  $\bar{M}_{\rm cl}$ is not large enough
for the initial clump  to open
a gap in the accretion disc. The newly-born clump will
therefore accrete from the disc. The Bondi-Hoyle 
estimate of the accretion
 rate gives $\dot{M}_{\rm cl}\sim \Omega \bar{M}_{\rm cl}$,
i.e. we expect the mass of the new clump to grow on the dynamical timescale
until it becomes large  enough to open a gap in the gas disc. The upper limit
$\tilde{M}_{\rm cl}$
of this value is the mass which opens a gap in the original gas disc with 
$\Sigma=\Sigma_{\rm crit}$ just
before it fragments:
\begin{equation}
\tilde{M}_{\rm cl}\simeq \bar{M}_{\rm cl}
\left[12\pi(\alpha_{\rm crit}/0.3)\right]^{1/2}(r/h_{\rm crit})^{1/2};
\label{tildeM}
\end{equation}
see Eq.~(4) of Lin and Papaloizou (1986).
Once the gas is depleted from the disc, we expect the initial distribution
of the clump masses  to be concentrated between $\bar{M}_{\rm cl}$ and $\tilde{M}_{\rm cl}$.
The clump masses will evolve when clumps begin to merge with each other; this
 is addressed in  section III.

\begin{figure}
\begin{center}
%\hbox{
\epsfig{file=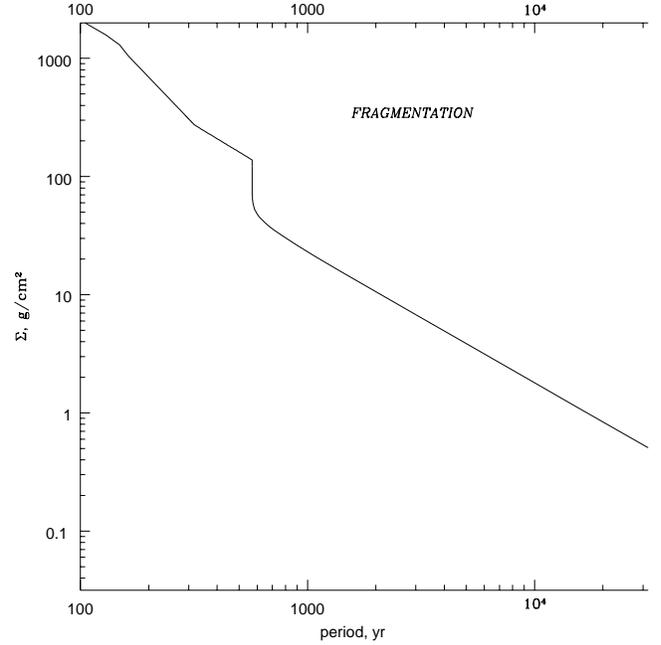, width=9cm}
%}
\end{center}
\caption{The surface density of the critically fragmenting disc as
a function of the orbital period.}
\end{figure}

\paragraph{Effect of irradiation and other sources of heating}. So far 
in determining the structure of the self-gravitating disc, we have 
neglected external or internal
 heating of the disc. This is certainly a poor approximation
in many cases. Irradiation from AGN or surrounding stars, or feedback
from star formation inside the disc can be the dominant source of heating
of the outer parts of AGN discs (eg, Shlosman and Begelman 1987, 1989).
For example, irradiation from circumnuclear stars will keep the disc
temperature at a few tens of Kelvin, which is higher than the critical
temperature we obtained for a self-gravitating disc beyond $0.1$pc 
away from $10^7M_{\odot}$ black hole; see our discussion for the Galactic Center
disc in section 4.  

The extra heating will always work to increase the critical temperature at
which the disc fragmentation occurs.
Therefore, the values of the  critical surface density $\Sigma_{\rm cr}$,  scaleheight $h_{\rm cr}$, 
and the mass of the initial fragment $M_{\rm cl}$ obtained above should be treated
as lower bounds of what might be expected around real AGNs or protostars.
Higher values of these quantities would only strengthen main conclusions
of this paper. 
LB have found that when the rate of accretion $\dot{M}$ is constant throughout the disc, the
Toomre mass $\bar{M}_{\rm cl}$ is given by
\begin{eqnarray}
\bar{M}_{\rm cl}&=&1.8M_{\odot}\left({\alpha\over 0.3}\right)^{-1}\times\nonumber\\
                & &{\dot{M}c^2\over L_{\rm edd}}
\left({M\over 3\times 10^6M_{\odot}}\right)^{0.5}\left({r\over 0.2\hbox{pc}}\right)
^{1.5},
\end{eqnarray}
and that the gap-opening mass is 
\begin{eqnarray}
M_{\rm gap}&=&62M_{\odot}\left({\alpha\over 0.3}\right)^{-0.5}\times\\
& &{\dot{M}c^2\over L_{\rm edd}}
\left({M\over 3\times 10^6M_{\odot}}\right)^{0.5}\left({r\over 0.2\hbox{pc}}\right)^{1.5}
\left({r\over 10h}\right)^{0.5}.\nonumber
\end{eqnarray}
Here, $M$ is the black-hole mass, and  $r$ and $h$ are the radius and the scaleheight
of the disc.
Thus, even if the clumps do not merge with each other
and stop their growth at the gap-opening mass, for high accretion rate the
mass of the formed stars will be biased towards the high-mass
end. In the next section we discuss the effects of clump mergers
and the mass range of stars born after the disc fragments.

\section{Evolution of the fragmented disc}

Gammie's simulations show that once the disc
fragments, the clumps merge and form significantly larger
objects. In fact, his razor-thin shearing box turned
into a single gas lump at the end of his simulation.

For merger to be possible,  the clumps should
not collapse into individual stars before they can
coalesce with each other. Let's check if this
is the case.

Consider a spherical non-rotating clump of radius $R$ and mass $M_{\rm cl}$.
First, assume that the clump is optically thin.
The  energy radiated  from the clump per unit time
is
\begin{equation}
W_{\rm cool}\sim \sigma T^4 R^2 \kappa(T)\Sigma\sim M_{\rm cl}\sigma T^4\kappa(T).
\label{Wcool}
\end{equation}
This radiated power cannot exceed the clumps gravitational binding energy
released in  free-fall time, $G^{1.5}M^{2.5}R^{-2.5}$. Together
with $\kappa(T)\propto T^2$ [since ice grains
dominate opacity for the optically-thin marginally
fragmenting disc--see Eq.~(\ref{tcrit1})],
this condition implies
that
\begin{equation}
T<\tilde{T}=T_0R^{-5/12},
\label{T1}
\end{equation}
where $T_0$ is a constant for the collapsing
optically thin clump. The temperature $\tilde{T}$ in
Eq.~(\ref{T1}) is less than the virial
temperature, which scales as $R^{-1}$.
Therefore, after the collapse commences, the clump is not virialized while
it is optically thin. The temperature cannot
be much smaller than $\tilde{T}$, since otherwise
the cooling rate would be much smaller than the rate
of release of the gravitational binding energy,
and the gas would heat up by quasi-adiabatic compression.
The inequality in Eq.~(\ref{T1}) should be
substituted
by an approximate equality, and therefore we have during
optically-thin collapse
\begin{equation}
T\propto R^{-5/12}.
\label{T2}
\end{equation}
The optical depth scales as
\begin{equation}
\tau\propto R^{-11/3}.
\label{tau2}
\end{equation}
and hence rises sharply as the
clump's radius decreases; as the clump shrinks
it  becomes optically
thick\footnote{The contraction of an optically thin clump
may be complicated by sub-fragmentation, since the
Jean's mass for such clump scales as $R^{7/8}\propto\tau^{-0.23}$.
We suspect that in most cases the clump becomes optically
thick before it sub-fragments, since the Jean's mass has a slow
dependence on the optical depth. However, only detailed numerical simulations
can resolve these issues.}. It is possible to show that once the clump
is optically thick, it virializes quickly with it's temperature
$T\propto R^{-1}$. For $\kappa\propto T^2$ (ice grains),
the cooling time of an optically thick clump scales with the clump radius as
\begin{equation}
t_{\rm cool}\propto R^{-3}.
\label{tcool}
\end{equation}
The characteristic timescale for the clump
to collide with another clump scales with the 
clump radius as 
\begin{equation}
t_{\rm collision}\propto R^{-2}.
\label{tcoll}
\end{equation}
From Eqs.~(\ref{tcool}) and (\ref{tcoll}),
we see that the collision rate decreases less
steeply
than the cooling rate as a function of the radius
of an optically thick clump. Therefore, merger
can be an efficient way of increasing the clump's mass.

This conclusion is no longer valid when the temperature
of the clump becomes  larger than $\sim 200$K; then
the opacity is dominated by metal dust with $\kappa\propto T^{1/2}$.
In this case the cooling time scales as $t_{\rm cool}\propto R^{-1.5}$.
The collision timescale increases faster than the cooling time
as the clump shrinks, and naively one would expect
that mergers may not be efficient in growing
the clump masses.

However,
 we have neglected  the rotational
support within a clump. Each clump is
initially rotating with angular frequency 
comparable to the clump's inverse dynamical timescale;
for example, in a Keplerian disc each clump's  
initial angular velocity is  $\sim \Omega/2$.
Therefore each clump will shrink and collapse
into a rotationally supported  disc, and the size
of this disc is comparable to the size of the original
clump (this picture seems to be in agreement with
Gammie's simulations).
Thus rotational support generally 
slows down the collapse of an individual fragment 
and makes mergers between different fragments
to be efficient.

 Magnetic braking is one of
the  ways for the clump to  lose its  rotational support\footnote{Another way
is via collisions with other clumps.}
(see, e.g., Spitzer 1978). 
One generally expects a horizontal magnetic field to be present
in a differentially rotating disc due
to the MRI (Balbus and Hawley, 1991). Ionization fraction
in the disc
is expected to be small,
so the magnetic field is saturated at a sub-equipartition
value $B=\beta B_{\rm eq}$, with $\beta<<1$. Horizontal magnetic field will couple 
inner and outer parts of the differentially rotating clump
on the Alfven crossing timescale $t_{\rm alfven}\sim t_{\rm dynamical}/\beta$,
and the collapse will proceed on this timescale as well.

\begin{figure}
\begin{center}
%\hbox{
\epsfig{file=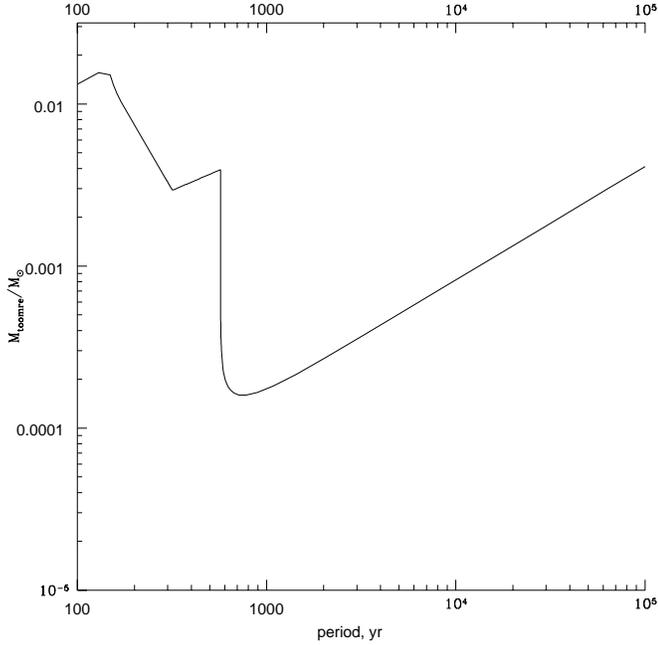, width=9cm}
%}
\end{center}
\caption{The Toomre mass of the critically fragmenting disc as
a function of the orbital period.}
\end{figure}

What is the maximum mass that the clump can achieve?
This issue has been analyzed for the similar situation
of a protoplanetary core accreting from a disc of
planetesimals (Rafikov 2001 and references therein).
The growing clump cannot accrete more mass than is 
present in it's ``feeding annulus''. This gives the maximum 
``isolation''  mass of a clump:
\begin{equation}
M_{\rm is}\sim {(2\pi r^2 \Sigma_{\rm crit})^{3/2}\over
         M^{1/2}}=2\pi\sqrt{2}\bar{M}_{\rm cl}(r/h_{\rm crit})^{3/2},
\label{Mis}
\end{equation}
where, as above, $\bar{M}_{\rm cl}=\Sigma_{\rm crit}h_{\rm crit}^2$
is the mass scale of the first clumps to form
from a disc; see, e.g.,  Eq.~(2) of Rafikov (2001). However, numerical
work of Ida and Makino (1993) and analytical calculations of
Rafikov (2001) indicate that the isolation mass may be hard
to reach. The consider a massive body moving on a circular orbit
through a disc of gravitationally interacting particles, and
they find that when the mass of the body exceeds some critical
value, an annular gap is opened in the particle disc around
the body's orbit.  We can idealize a disc consisting of fragments
as a disc of particles of a typical fragment mass $M_{\rm fr}$.
Once a growing clump opens a gap in a disc of gravitationally
interacting fragments, the clump's growth may become quenched.
 This gap-opening  mass of the
clump $M_{\rm gap}$  is given by Eq.~(25) of
Rafikov (2001):
\begin{equation}
{M_{\rm gap}\over M_{\rm is}}={I\over 2^{7/6}\pi^{1/2}}
\left({M_{\rm fr}\over \Sigma r^2}\right)^{1/3} \left(
{M\over \Sigma r^2}\right)^{1/2}.
\label{rafikov25}
\end{equation}

 We use the numerical factor
$2^{-7/6}\pi^{-1/2}I=1.5$ appropriate for thin discs. 
By taking $Q=1$ we get 
\begin{equation}
M_{\rm gap}\simeq 14 M_{\rm fr} (M_{\rm fr}/\bar{M}_{\rm cl})^{1/3}(r/h_{\rm crit}).
\label{mgap}
\end{equation}
In Figure 4  the masses $M_{\rm is}$
and $M_{\rm gap}$ are plotted
as a function of radius for a $3\times 10^6M_{\odot}$
black hole; when we calculate  $M_{\rm gap}$ we
conservatively set $M_{\rm fr}=\bar{M}_{\rm cl}$ and not
to the larger value $\tilde{M}_{\rm cl}$. It is likely that the most massive 
clumps will reach $M_{\rm gap}$,
but it will  be more difficult to form a clump
with the mass $M_{\rm is}$.

From Fig.~$4$ we see that the most 
massive clumps can reach tens hundreds of solar masses.
The expected e-folding time for the clump growth is
comparable to the orbital period, and thus these high masses
will be reached within $\sim 20$ orbital periods.
The maximum mass would be even larger if we included
the heating of the disc by external irradiation or 
internal starburst.
It is plausible that these  
 very massive clumps will form   massive stars;
the masses of the stars may be comparable to the
masses of the original clumps; see McKee and Tan (2002)
and references therein.

Goodman and Tan (2004) have argued that a protostar
accreting from the accretion disc will reach the isolation mass;
see also Nayakshin and Cuadra (2005) and Nayakshin (2005).
This conclusion was criticized by Milosavljevic and Loeb (ML, 2004),
who have shown that the mini-accretion disc around the star itself will
be unstable to fragmentation. Our picture of merging clumps forming a massive
protostellar core is not susceptible to the ML's criticism.

\section{Young stars near SgrA*}
\subsection{Fragmenting disc}
So far we have mostly considered an idealized situation where the disc is self-luminous
and not irradiated externally. However, a gaseous disc in the Galactic Center (when there is one)
%a disc of say $M_{\rm disc}=10^4M_{\odot}$ in the GC;
%this mass is of the order of magnitude needed to form the early-type
%stars.
%The column density of the gas is then
%\begin{equation}
%\Sigma\sim {M_{\rm disc}\over \pi r^2}\simeq 0.65 \left({M_{\rm disc}\over
%10^4M_{\odot}}\right)(r/\hbox{pc})^{-2},
%\label{sigma}
%\end{equation}
%where $r$ is the radial extent of the disc. 
is heated
from outside by the light from bright stars in the cusp (mostly UV),
 and  re-radiates this in the
infrared. If the disc is  optically thick to its thermal radiation,
then the temperature $T$ of the disc is uniform in vertical direction
and is determined by the Stefan-Boltzmann law:
\begin{equation}
\sigma T^4\sim {L_{\rm irradiation}\over \pi R^2}.
\label{T1} 
\end{equation}
where $R$ is the disc radius.
If the disc is  optically thin to its thermal radiation,
 then determining the  vertical temperature profile of the disc requires more
care.
The superheated dust layer forms on its skin and the interior is
heated by the infrared light
emitted by this layer (Chiang and Goldreich, 1997).
 Two cases can be distinguished:
first, in which the disc interior is thin to its own thermal radiation
but thick to the one from superheated dust, and second, in
which the disc is thin to both its own radiation and that
of the dust. In both
cases, however, the mid-plane temperature will be slightly, but
not much, greater then the estimate in Eq.~(\ref{T1}) (i.e.,
by a factor of order $\tau^{-1/4}$ in the first case and
$(\epsilon_s/\epsilon_i)^{1/4}$ in the second case; here
$\tau$ is the disc's optical depth to its thermal radiation
and $\epsilon_{s,i}$ are the dust emissivities in the superheated layer
and mid-plane, respectively). We will see that the actual optical depth of the
disc to its thermal radiation will be of order 1.

The current total luminosity of stars is estimated to be 
$L_0=2\times 10^7 L_{\odot}$ (Paumard et al., 2005). Lets assume that a fraction $\beta$ 
of the stellar light
is intercepted by the disc (this fraction is of order 1 since the disc is optically
thick for the UV). From Eq.~(\ref{T1}), the disc temperature
is\footnote{The disc will be also heated by the stars from the central star cluster 
when they collide with the gas (Syer et al., 1981). The rate of heating per area can be estimated as
 $\sim \Sigma r_*^2 n \sigma^3\simeq 5L_{\odot}/\hbox{pc}^2$, where $\Sigma$ is the disc column density, $r_*$ is the typical stellar radius,
and $n$ and $\sigma$ are the density of stars and the velocity dispersion of the cluster, respectively. In the numerical estimate
we have taken the values appropriate for the Galactic Center environment:
$r=r_\odot$, $\Sigma=10\hbox{g}/\hbox{cm}^2$, $n=10^6\hbox{pc}^{-3}$, and $\sigma=1000$km/s. Clearly, the power input from this process is negligibly
small compared to that from the external irradiation.} 
\begin{equation}
T\simeq 84\beta^{1/4}(L/L_0)^{1/4}(0.5\hbox{pc}/R)^{1/2}\hbox{K}.
\label{T2}
\end{equation}
and the speed of sound is
\begin{equation}
c_s\simeq 4.5\times 10^4 \beta^{1/8}(L/L_0)^{1/8}(0.5\hbox{pc}/R)^{1/4}\hbox{cm/sec}
\label{cs}
\end{equation}

From Figure 1 we see that for orbital periods greater than 600 years (this
corresponds to $r>0.06$pc for SgrA*) the temperature of self-luminous disc is
smaller than that of the gaseous disc in the Galactic Center, and thus the disc
is supported by external irradiation prior to fragmentation. At the fragmentation
boundary,
the Toomre $Q$ parameter equals 1:

\begin{equation}
Q={c_s\Omega\over \pi G \Sigma}=1
\label{Qcs}
\end{equation}
and thus from Eq.~(\ref{cs}),
\begin{equation}
\Sigma=2.7\times(0.5\hbox{pc}/r)^{-3/2}(\beta L/L_0)^{1/8}
(0.5\hbox{pc}/R)^{1/4}\hbox{g}/\hbox{cm}^2
\label{sigmasgr}
\end{equation}
Thus the {\it mass} of the fragmenting disc is 
\begin{eqnarray}
M_{\rm disc}&=&4\pi (R^2*\Sigma(R)-R_{\rm in}^2\Sigma(R_{\rm in}))\nonumber\\
            &=&2.5\times 10^4 [(R/0.5\hbox{pc})^{1/2}-(R_{\rm in}/0.06\hbox{pc})^{1/2}]\times\\	              & &(\alpha L/L_0)^{1/8}(0.5\hbox{pc}/R)^{1/4}M_{\odot},\label{Mdisc}
\end{eqnarray}
where $R_{\rm in}$ is the inner radius of the disc. I have taken this inner radius to
be $R_{\rm in}=0.06$pc, since inside this radius the critical column density necessary
to achieve fragmentation is much larger than outside, see Fig.~2. The value of
$R_{\rm in}$ agrees well with the observed inner edge of the clockwise stellar disc, which
is $0.03$pc in projection on the sky (Paumard et al., 2005). The disc mass obtained
 in the Equation above appears to be twice as large as the value estimated from
observations by Paumard et al.~(2005); see also Nayakshin et al.~(2005). 
The difference can be explained if either (a)
the stellar luminosity was smaller by $\sim 300$ than the current one during the
disc formation; this is unlikely since even though most of the current luminosity is due to 
the recently-formed stars, there is
substantial UV flux from post-AGB stars in older populations and the
center of the galaxy has significant soft-x-ray sources which should
predate the most recent starburst,
(b) the starburst occupied the outer half of the disc,
or (c) only half of the gas was converted into stars, and the rest left the GC as wind.

Equation (\ref{sigmasgr}) makes a prediction that $\Sigma\propto r^{-1.5}$ during fragmentation.
But what about the number density of formed stars? If the typical stellar mass scales with the
isolation mass of the disc (Goodman and Tan 2004, section 3 of this paper), then

\begin{equation}
M_{\rm isolation}\propto (\Sigma r^2)^{3/2}\propto r^{0.75},
\end{equation}
and the number of stars per unit are scales as
\begin{equation}
dN/d(Area)\propto r^{-2.25}.
\end{equation}
This is in good agreement with the data in Paumard et al.~(2005), who
find $dN/d(Area)\propto r^{-2}$

\begin{figure}
\begin{center}
%\hbox{
\epsfig{file=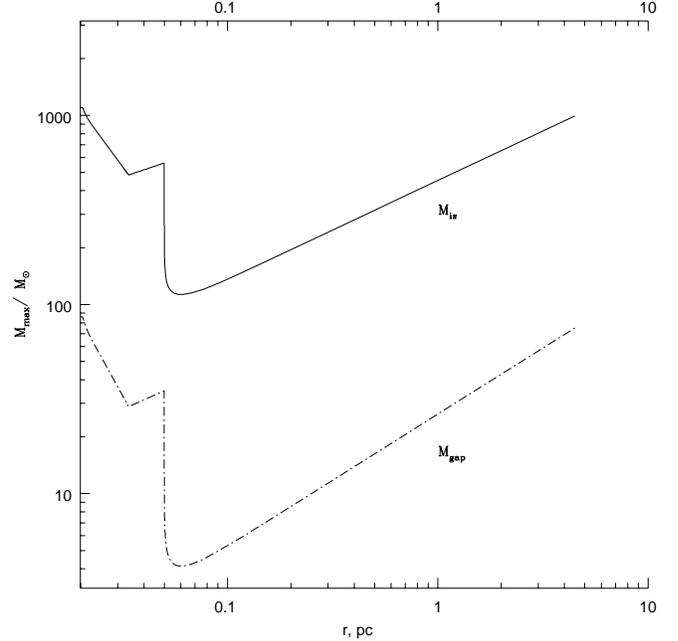,width=9cm}
%}
\end{center}
\caption{The isolation and gap opening masses plotted
as a function of radius for the critically fragmenting disc without
external sources of heating. The black hole mass is taken to be
$3\times 10^6M_{\odot}$.}
\end{figure}

\subsection{S-stars}
It is not easy to visualize how eccentric orbits orbits of the S-stars
could be consistent with their birth in a disc, and so far the ideas for
their origin have invoked stellar dynamics.
Recently, two interesting suggestions have been put forward. 
Alexander and Livio (2004) have proposed that the S-stars got captured by
an exchange interaction with stellar-mass black holes in the SgrA* cusp; however
this scenario seems to be disfavored by current observations (Paumard et al, 2005).
Gould and Quillen (2003) have argued that in the 
past the S-stars were members of binaries. When the binary passes close to
SgrA*, one of its members gets ejected and the other remains bound to
SgrA* on an eccentric orbit. However, (a) it is unclear how a young binary would get
onto a plunging orbit during its lifetime, and (b) some of the S-stars' eccentricities 
are not high; for example S1 and S13 have eccentricities of around 0.4 (Eisenhauer et al., 2005).
 It is certainly
impossible to capture stars on orbits with such low eccentricity by disruption of 
stellar binaries, since the typical eccentricity scales as $(1-e)\sim (m/M)^{1/3}$ where
$m$ and $M$ is the mass of the binary and SgrA*, respectively. Therefore, one has to appeal to a fast relaxation
twice: the first time, to get the binary on a plunging orbit when it is far from SgrA*, and the second time,
to get the star's eccentricity relaxed to a lower value when it is close to SgrA*. 

Rauch and Tremaine  (1996)
have identified a fast relaxation process near supermassive black holes, and called it the ``resonant relaxation''
(RR). Here we argue that it is potentially important for the S-stars; this argument has also been
made independently by Hopman and Alexander (2006)\footnote{We acknowledge Scott Tremaine 
for attracting our attention to the RR as a potentially important relaxation mechanism for
the S-star relaxation, in 2004.} The characteristic timescale for the eccentricity change due to RR is given by
\begin{equation}
t_{RR}\simeq (0.5/\beta_s)^2\left({M\over 4\times 10^6 M_{\odot}}\right)
             \left({16M_{\odot}\over m_s}\right)(P_{\rm orb}/\hbox{yr})\times 10^6\hbox{yr},
\label{trr}
\end{equation}
where $m_s$ is the typical mass of a star in the cusp [which is thought to be dominated
by stellar-mass black holes in the central $0.1$pc, see Miralda-Escude and Gould (2000)], 
$P_{\rm orb}$ is the orbital period of the star, and $\beta_s$ is the numerical coefficient,
found to be approximately $0.5$ by Rauch and Tremaine. Equation (\ref{trr}) is valid so long
as relativistic precession is longer than the Newtonian precession due to the cusp. Otherwise,
the expression for $t_{RR}$ has to be multiplied by the ratio of the two precession frequencies.
The longest measured period of the 
S-star is that of S1, which has $P_{\rm orb}=94$yr (Eisenhauer et al., 2005, see also 
Ghez et al., 2005). Thus the longest RR timescale for the
S-star is of order $100$Myr, well within a possible lifetime of the main sequence B stars.
The RR relaxation timescales are shorter for all the other stars, the shortest one being for
S2: $t_{RR}\simeq 40$Myr. Given that we do not know the typical mass of a stellar-mass black hole
in the cusp\footnote{Top-heavy IMF may imply top-heavy black holes!}, we conclude that 
{\bf RR is a viable
mechanism for relaxation of orbits of all of the S-stars within their lifetime}.

However, there is no mechanism known to us which would place the young stars on  plunging orbits when
they are 
further out from the SgrA*, as required in both Gould and Quillen (2003) and Alexander and Livio (2004)
scenarios. We therefore suggest that the S-stars were born during the previous starburst or several
starbursts, and then migrated towards smaller radii through their parent discs. This possibility
was already pointed out in LB; it is known as type-I migration in planetary dynamics.
The migration velocity through the disc is given by  
\begin{equation}
v_{\rm in}\sim 2\beta {G^2m\Sigma h\over c_s^3r};
\label{typeIv}
\end{equation}
cf.~Eqs.~(9), (B4) and (B5) of Rafikov (2002), and Ward (1986).
Here $\beta$ is a numerical factor of order $5$ for  $Q\gg 1$, 
and can be significantly larger for $Q\sim 1$. 
From Eq.~(\ref{cs}) we see that the migration timescale for an S-star embedded
in the disc at $0.1$pc is of order $10^5$yr, i.e.~much shorter than the star's lifetime.
Complication arises if the star opens a gap in the disc.
However, in this case a fast runaway migration is possible due to co-rotation
torques acting on the star; the speed of the runaway migration is comparable
to that of type-I migration (Artymowicz 2003, Masset and Papaloizou 2003).

\section{Merger of the central black hole and the disc-born black hole.}
 Stars with masses of a few tens
of solar masses 
will produce black holes as the end product
of their rapid ($\sim 10^6$yr) evolution; the characteristic
mass of these black holes is believed 
to be around $10M_{\odot}$. A recent work by Mirabel and
Rodrigues (2003)
shows that the  black holes whose progenitors have
masses $>40M_{\odot}$ do not receive a velocity kick at their birth.
Thus, significant fraction of the disc-born black holes will remain embedded
in the disc.

It is likely that the newly-born black hole
in the disc  inspirals towards the central
black hole. We imagine that 
 the stellar-mass black hole
is embedded into a massive accretion disc which
forms due to continuing infall of gas from the
galactic bulge, after the black hole is born. If the
black hole opens a gap in the disc, it will move
towards the central black hole together with the disc
(type-II migration; Gould and Rix 2000, and Armitage
and Natarajan 2001). The timescale for such inspiral
is the accretion timescale,
\begin{equation}
t_{\rm inspiral}\sim 
10^6\hbox{yr}{M_7^{-1/2}\over \alpha_{0.1}}\left(
{0.1\hbox{pc}\over r}\right)^{-3/2}\left({r\over 30h}\right)^2.
\label{typeII}
\end{equation}
If on the other hand, the black hole is not massive 
enough to open the gap, it will migrate inwards by
exciting  density waves in the disc (type-I migration, see previous section).
The speed of this inward drift is given by
Eq.~(\ref{typeIv}).
For a self-gravitating
disc with $Q\sim 1$, we find    
\begin{equation}
t_{\rm inspiral}\sim 10^7\hbox{yr}{5\over\beta}{30h\over r}{10M_{\odot}
\over M_{\rm bh}}\left({r\over 0.1\hbox{pc}}\right)^{1.5}M_7^{0.5}.
\label{typeI}
\end{equation}

What determines whether the gap is open or not is whether a 
stellar-mass black hole has time to accrete a few thousand solar
masses of gas, which would put it above the gap-opening threshold
for the typical disc parameters. The Eddington-limited accretion
occurs on a timescale of $\sim 10^8$yr, much longer than the
characteristic type-I inspiral time. Thus, in this case the black
holes don't accrete much on their way in.
On the other hand, the Bondi-Hoyle formula predicts the mass e-folding
timescale of a few hundred years. Thus, if the black hole
is allowed to accrete at the Bondi-Hoyle rate, its mass
increases rapidly until it opens a gap in the disc. Then, the
inspiral proceeds via type-II migration.

From Eqs.~(\ref{typeII}) and (\ref{typeI}) we see that the embedded
black hole experiencing type-I or type-II migration
would merge with the central black hole on the timescale $10^6$---$10^7$
years, shorter than the typical timescale of an AGN activity.
Thus it is plausible that the daughter  disc-born
black hole is 
brought towards the parent central black hole; the mass of the
daughter black hole  might grow significantly on the way in. Gravitational radiation
will eventually become the dominant mechanism driving the inspiral,
and the final merger will produce  copious amount of gravitational
waves. In the next subsection we show that
these waves  are detectable by LISA for a broad range of
the black hole masses and the disc accretion rate.

\subsection{Influence of the  accretion disc on the inspiral signal 
as seen by LISA}

It is realistic to expect that LISA would follow the last year of the inspiral
of the disc-born black hole into the central black hole.
Generally, one must develop a set of templates which
densely span the parameter space of possible inspiral signals. In order
for the final inspiral to be detectable, one of the
templates must follow the signal with the phase-shift between the two not exceeding a fraction
of a cycle. Therefore, if the drag from the disc
will alter the waveform by a fraction of a cycle over the signal integration
time (e.g., $1$ year),   detection of the signal  with high
signal-to-noise ratio will become problematic.
Below we address the  influence of the accretion disc  on
the final inpiral waveform.

The issue of gas-drag influence on the LISA signal was first
addressed by Narayan (2000); see also 
Chakrabarti (1996). Narayan's analysis is directly applicable
to low-luminosity non-radiative quasi-spherical accretion flows,
which might exist around supermassive black holes when the accretion rate
is $<0.01$ of the Eddington limit. Narayan concluded that non-radiative
flows will not have any observable influence on the gravitational-wave signals
 seen by LISA. Below we extend Narayan's analysis to the case of
radiative disc-like flows with high accretion rate, which are likely
to be present in high-luminosity AGNs.

Consider a non-rotating central black hole of mass $M_6\times 10^6M_{\odot}$,
accreting at a significant fraction $\dot{m}$ of the
Eddington rate, $\dot{M}=\dot{m}\dot{M}_{\rm edd}$.
The accretion disc in the region of interest ($<10 R_S$, where
$R_s$ is the Schwartzschild radius of the central black hole) is radiation-pressure
dominated, and the opacity $\kappa=0.4\hbox{cm}^2/\hbox{g}$ is due
to the Thompson scattering. By following the standard thin-disc 
theory\footnote{The disc is no longer thin close
to the central black hole, however our estimates of
the disc structure should be correct to an order of magnitude.}
(Shakura and Sunyaev, 1973), we get
\begin{equation}
\Sigma(r)={64\pi c^2\over 27\alpha\Omega\kappa^2\dot{M}}\sim 
4\hbox{g}/\hbox{cm}^2{\epsilon\over \alpha \dot{m}}\left(
        {r\over r_s}\right)^{3/2},
\label{sigmabh}
\end{equation}
where $\epsilon$ is the efficiency with which the accreted mass converts into
radiation, and $r_s$ is the Schwartzschild radius of the central black hole;
and 
\begin{equation}
c_s={3\dot{M}\Omega\kappa\over 8\pi c}.
\label{csbh}
\end{equation}

The disc black hole in orbit around the
central black hole  excites density
waves in the disc (Goldreich and Tremaine, 1980); these waves carry angular momentum flux
\begin{equation}
F_0\sim \left(GM_{\rm bh}\right)^2 {\Sigma r \Omega\over c_s^3};
\label{fluxwaves}
\end{equation}
here, as before,  $M_{\rm bh}$ is the mass of the orbiting disc black hole.
Ward (1987) has argued that the torque acting on the
orbiting body is $dL_{\rm dw}/dt \sim (h/r)F_0$. We can compute the
characteristic timescale for the orbit evolution due to the 
density-waves torque:
\begin{equation}
t_{\rm dw} = {L\over dL_{\rm dw}/dt}={1\over \Omega}{M\over M_{\rm bh}}{M\over \Sigma r^2}\left({h\over r}\right)^2.
\label{torquewaves}
\end{equation}
One must compare $t_{\rm dw}$
to the timescale $t_{\rm gw}$ of orbital evolution due to gravitational-radiation-reaction torque:
\begin{equation}
t_{\rm gw}=8t_{\rm m}={5\over 8}{cr_s^2\over GM_{\rm bh}}\left({r\over r_s}\right)^4.
\label{tgw}
\end{equation}
Here, $t_m$ is the time left before the disc and central holes merge, i.e.~the
integration time for the LISA signal.
Optimistically we could expect to follow the LISA signal to $0.1$ of a cycle.
 Therefore, if $q=10 n_{\rm m} t_{\rm gw}/t_{\rm dw}$ is less than
unity, the disc drag does not impact detection of the final inspiral; cf.~Eq.~(16)
of Narayan (2000). Here $n_{\rm m}=\Omega t_{\rm gw}/(5\pi)$ is the number
of cycles the disc black hole will make before merging with the central black hole.
Using Eqs.~(\ref{sigmabh}), (\ref{csbh}), (\ref{torquewaves}), and (\ref{tgw}), we 
get
\begin{equation}
q\simeq 2\times 10^{-7}{\epsilon_{0.1}^3\over \dot{m}^3 \alpha_{0.1}} 
{(M_{\rm bh}/10M_{\odot})^{13/8}\over M_6^{13/4}}
t_{m, yr}^{21/8},
\label{q}
\end{equation}
where $\epsilon=0.1\epsilon_{0.1}$, $\alpha=0.1\alpha_{0.1}$, and
$t_m=1\hbox{yr}*t_{m, yr}$.
We see that for a large range of parameters $q<1$, and the disc drag does not
influence the inpiral signal. However, note that $q$ is a very sensitive
function of $\dot{m}$ and $M_{\rm bh}$. For instance, the disc will influence significantly
an inspiraling $100M_{\odot}$ black hole when the accretion rate is down to a few
percent of the Eddington limit. One then needs to reduce the influence
of the disc on the LISA signal
 by choosing to observe the smaller portion of the final
inspiral, i.e.~by choosing a smaller integration time $t_m$.

There is another important source of drag experienced by the inspiraling
black hole;  it was first analyzed by Chakrabarti (1993, 1996). Generally,
there is a radial pressure gradient in an accretion disc; this pressure 
gradient makes the azimuthal velocity of the disc gas  slightly different
from a   velocity of the test particle on a circular orbit at the same radius.
Therefore the  inspiraling black hole will experience a 
head wind from a gas in the accretion disc;  by accreting gas
from the disc the black hole will experience  the braking torque which
will make it lose it's specific angular momentum. This torque is given 
by 
\begin{equation}
\tau_{\rm wind}=\dot{M}_{\rm bh}\Delta v r,
\label{tauwind}
\end{equation}
where $\dot{M}_{\rm bh}$ is rate of accretion onto the inspiraling 
black hole from the disc, and $\Delta v\sim c_s^2/v_o$ is the speed of the 
headwind experienced by the black hole moving with the orbital speed $v_o$.
We assume that the inspiraling hole accretes with the Bondi-Hoyle rate,
\begin{equation}
\dot{M}_{\rm bh}\simeq \pi\rho (GM_{\rm bh})^2/c_s^3,
\label{mdisc}
\end{equation}
where $\rho=\Sigma/h=\Sigma\Omega/c_s$ is the density of the ambient
disc gas.  By using the last equation in Eq.~(\ref{tauwind}), we obtain
the expression for the torque $\tau_{\rm wind}$, which turns out
to be the same as the torque from the density waves, up to the numerical
factor between $1$ and $10$. We see therefore that inclusion of Chakrabarti's
``accretion'' torque is important  for the detailed analysis,
but does not qualitatively change our conclusions.    

Can the orbiting  black hole open  
a gap in the disc during it's final inspiral?
In order to overcome the viscous stresses which oppose  opening
the gap, the mass of the orbiting black hole should be greater than
the  threshold value
given by 
\begin{equation}
M_{\rm bhgap}\simeq\sqrt{40\alpha}(h/r)^{2.5}M,
\label{mbhgap}
\end{equation}
see Eqs.~(4) of Lin and Papaloizou (1986).
The radius $r_{\rm in}$ from which the inspiral begins is 
given by 
\begin{equation}
r_{\rm in}=4(M_{\rm bh}/10M_{\odot})^{1/4}M_6^{-1/2}t_{m, yr}^{1/4}r_s,
\label{rin}
\end{equation}
and the scaleheight of the disc is radius-independent in the radiation-dominated
inner region (this is true only if you treat the central black hole as a Newtonian
object):
\begin{equation}
h={3\dot{m}\over 4\epsilon}r_s.
\label{hin}
\end{equation}
The gap will be open in the disc, therefore, if the mass of
the inspiraling black hole exceeds
\begin{equation}
m_{\rm gap}\sim \dot{m}^{20/13}M_6^{-2/13} 10^4M_{\odot}.
\label{mvgap1}
\end{equation}
During the final year of merger of the   $10M_{\odot}$ disc-born black hole
and  the $10^6M_{\rm odot}$ central black hole, the gap will
be open if the accretion rate onto the central
black hole is about five  percent of the Eddington limit.
Such gap-opening merger can result in sudden changes in the
AGN luminosity and produce an optical counterpart to the 
gravitational-wave signal. 

\subsection{The merger event rate as seen by LISA}
The details of black-hole formation and evolution in the disc are 
uncertain, and   it seems impossible to make a reliable estimate
of an event rate for LISA from this channel of black-hole mergers.
Nonetheless, the expected top-heavy IMF gives  us reason for optimism. 
It is worth working through a simple example to
show that the merger of disc-born holes with the central hole
may  be an important source for LISA.

Studies of integrated light coming from Galactic Nuclei show
that the supermassive black holes acquire significant part and
perhaps almost all of their mass via accretion of gas;
see, e.g., Yu and Tremaine (2002) and references therein. Assume,
for the sake of our example, that a mass fraction $\eta$ of this gas
is converted into $100M_{\odot}$ black holes on the way in,
and that all of these black holes eventually merge  with the central
black hole. LISA can detect such mergers to $z=1$ if the central
black hole is between $10^5$ and $10^7$ solar masses (one needs to assume
that the central black hole is rapidly rotating at the high-mass end
of this range). The mass density of such black holes in the
local universe is $\sim 10^5 M_{\odot}/\hbox{Mpc}^3$ (Salucci et.~al.~1999).
We can estimate, using Figure $3$ of Pei et.~al.~(1995),
that about $10$ percent of  integrated radiation from Galactic Nuclei
comes from redshifts accessible to LISA, $z<1$. This implies that
 supermassive black holes acquired $10$ percent of their mass
at $z<1$. We shall therefore assume\footnote{This assumption has an extra caveat for the
low-luminosity AGNs, significant fraction of which may be powered by tidal disruption
of stars (Eracleous et al.~1995, Milosavljevic et al.~2006). We note that in one low-luminosity
AGN, in the nucleus of Circinus galaxy, there is a firm evidence of 
an extended accretion disc (Greenhill et al.~2003).} that $10$ percent of the mass of the
black holes in LISA mass range was accreted at $z<1$. This implies that
there were $\sim 100\eta=\eta_{0.01}$ mergers per mpc$^3$ which are potentially detectable
by LISA. When we multiply this by volume out to $z=1$ and divide it by
the Hubble time, we get an estimate of the LISA event rate from 
such mergers,
\begin{equation}
dN/dt_{\rm toymodel}\sim 10\eta_{0.01}/\hbox{yr}.
\label{rate}
\end{equation}
The estimate above should be treated as an illustration of importance
for our channel of the mergers, rather than as a concrete prediction for LISA.

Currently, it is not
known how to compute well the gravitational waveform produced
by an inspiral with an arbitrary eccentricity and inclination relative to the central 
black hole. This might pose a great challenge to the LISA data analysis.
Indeed, in the currently popular  astrophysical scenario, 
the stellar-mass black hole gets captured on a highly eccentric
orbit with arbitrary inclination relative to the
 central black hole (Sigurdsson and Rees, 1997).
By contrast, in our scenario the inspiral occurs in the equatorial plane, and the signal is well
understood (Hughes 2001 and references therein). Circular inspiral is also expected in the
binary-disruption scenario of Miller et al.~2005; however there the inspiral is not
expected to be preferentially aligned with the equatorial plane. The
template for detection of a circular inspiral is readily available, and
the merger channel we consider has a clear observational
signature which distinguishes
it from other channels. It is an open question whether the inspiraling
hole can acquire high eccentricity by interacting with the disc or with other orbiting masses
(Goldreich and Sari 2003, Chiang et.~al.~2002); it seems likely that at least in some
cases interaction with the disc will act to circularize the orbit. Finn and Thorne (2000) have
performed a detailed census of parameter space for circular-orbit equatorial  inspirals as seen by LISA.  

\section{conclusions}
In this paper,
we rely on  numerical simulations by Gammie (2001)
to develop a  formalism for self-gravitating 
thin discs which are gaining mass by continuous infall.
We present a  way to calculate the critical
temperature, surface density, and scaleheigh of the
disc just prior to fragmentation. Our formalism naturally 
includes both 
optically thin and optically thick discs.

We then speculate on the outcome of the nonlinear physical
processes which follow fragmentation: accretion and merger
of smaller fragments into bigger ones. We find an upper
bound on the mass of final, merged fragments, and we give
a plausibility argument that some fragments will
indeed reach this upper bound. We thus predict that
very massive stars of tens or even hundreds of solar masses
will be produced in self-gravitating discs around supermassive
black holes. This picture explains well recent observations
of the stellar discs
in the Galactic Center, and we clarify how the 
peculiar orbits of the S-stars may be generated by a combination
of inward migration through the gaseous disc and Resonant Relaxation
in the central arc-second.

The end product of fast evolution of the massive
stars will be stellar-mass black holes. 
We make an  argument that in the case when accretion is 
prolonged, 
the disc-born black holes in AGNs  find a way to merge
with  central black holes. We consider a purely toy-model
example of what the rate of such mergers might be, as seen
by LISA; we illustrate this rate might  be high enough 
to be  interesting 
for future gravitational-wave (GW) observations.
The  GW signal from this merger channel is
distinct from that of other channels, and can be readily
modeled using our current theoretical understanding of the 
final stages of the inspiral driven by gravitational-radiation reaction.
We 
 show that for a broad range
of accretion rates and black-hole masses
the drag from accretion disc will not be large enough
to pollute the signal and make inspiral template invalid.
In some cases, the inspiraling
hole will open a gap in the accretion disc close to  
the central black hole, thus producing  a possible optical
counterpart for the gravitational-wave burst generated    
by  the merger. 

\section{acknowledgments}
I have greatly benefited from discussions
with Chris Matzner and Scott Tremaine.
This paper is largely based on my earlier
preprint (Levin, 2003).
I thank Norm Murray and Clovis Hopman for encouraging
me to publish it.

\end{document}